\documentclass{PoS}
\usepackage{amsmath,amssymb}
\usepackage{mcite}
\newcommand{\Tr}[0]{\text{Tr}}
\newcommand{\sign}[0]{\text{sign}}
\newcommand{\path}[1]{ #1}
\title{Multi-grid HMC for Ginsparg-Wilson fermions}

\ShortTitle{Milti-grid HMC}

\author{\speaker{Nigel Cundy}\\
        Universit\"at Regensburg, Universit\"at Strasse 31, 93040 Regensburg, Germany\\
        E-mail: \email{nigel.cundy@physik.uni-regensburg.de}}

\abstract{I describe a method that places the fermion fields and the gauge fields on different lattice spacings during the Hybrid Monte Carlo generation of Ginsparg-Wilson dynamical ensembles. The idea is motivated by Wilson's formulation of the renormalisation group. After outlining the underlying theory, I describe a method to perform most of the work of the HMC on a coarse lattice, only requiring to convert to and from the fine lattice once for each independent configuration. Because the bulk of the work takes place on the coarse lattice, including the calculation of most observables, this method saves over an order of magnitude in computer time over previous methods.}

\FullConference{The XXVII International Symposium on Lattice Field Theory\\
                 July 26-31, 2009\\
                 Peking University, Beijing, China}

\begin{document}

\section{Introduction}
Overlap fermions~\cite{Narayanan:1993ss,*Narayanan:1993sk,*Neuberger:1998my} are unquestionably the theoretically cleanest formulation of lattice QCD, because they alone exactly fulfil the Ginsparg-Wilson relation~\cite{Ginsparg:1982bj} and satisfy an exact chiral symmetry on the lattice~\cite{Luscher:1998pqa}. However, they are also unquestionably the most expensive formulation. 

In~\cite{Cundy:2009ab}, I explored the possibility that the lattice overlap Dirac operator is connected to the continuum by a block renormalisation group transformation; this was motivated by consideration of  their obedience to the renormalisation group derived Ginsparg-Wilson relation. The construction contained within that paper was incomplete, because it did not address how to construct the Yang-Mills action in a similar way. I will finish this work in a future paper by using a coordinate transformation of the gauge fields, and I briefly outline the method in section \ref{sec:2.3} of this work.  Once the suitable transformation is constructed, and the correct form of the Yang-Mills action found, then it will be possible to transfer from one lattice spacing to another, including the continuum,  through the renormalisation group.

This opens up the possibility of running lattice simulations at two length scales: a fine lattice for the gauge fields, and a coarse lattice for the fermions. The connection between the coarse lattice and an equivalent action on the fine lattice will be provided by renormalisation. The advantages of this approach are obvious: if we use a scale factor of $s$ between the fine and coarse lattices, then we save at least a factor of $s^4$ in computer time for the expensive parts of the calculation. However, the physics would be identical on the coarse lattice and the fine lattice, which itself can be extracted if necessary at an additional cost, but will only be required for certain observables which require a finer resolution (for example in studies using heavy quarks). The challenge is linking the coarse action with a fine action. One possibility is re-weighting, but the re-weighting coefficients will  be of order $e^{\alpha_6 a^6 V + \ldots}$ (the leading order $a^4V$ term can be absorbed into the Yang-Mills action), which are likely to be unstable at larger volumes without fine tuning of $\alpha_6$ and the terms hidden by the ellipsis. A better approach is to find some method of interpolating between the coarse action and the fine action; and here I will present one possible way of doing this.   

These proceedings are intended only as a very preliminary description of the proposed method. A full proof of the method's validity, numerical results, and a discussion of whether this is indeed beneficial will follow in a subsequent paper. In section \ref{sec:2}, I outline the underlying renormalisation group theory. In section \ref{sec:3}, I outline the proposed Monte-Carlo routine; In section \ref{sec:4}, I describe the numerical implementation which I will use in my tests, and, in section \ref{sec:5}, I conclude.

\section{Renormalisation group and overlap fermions}\label{sec:2}
\subsection{Gauge Action}
The multi-grid HMC algorithm presented in section \ref{sec:3} requires that the action is written entirely in terms of the Dirac operator (I shall justify this statement in the follow-up paper). This includes the Yang-Mills term. I note that in the continuum, if $D = \gamma_{\mu} (\partial _{\mu} - A_{\mu})$, then
\begin{gather}
-i\sigma_{\mu\nu}F_{\mu\nu} = (\gamma_5D)^2 - \frac{1}{4}\Tr_S (\gamma_5D)^2,
\end{gather}
where the trace is only over the spinor indices. Therefore, the Yang-Mills action can be written as
\begin{gather}
S_g = \frac{\beta}{4}\Tr F_{\mu\nu}^2 = \frac{\beta}{4} \Tr[(\gamma_5D)^4] - \frac{1}{4}\Tr[(\gamma_5D)^2\Tr_S[(\gamma_5D)^2]].  
\end{gather}
This can obviously and easily be extended to any lattice formulation of the Dirac operator, although for any Dirac operator which is not ultra local, including the overlap, the numerical implementation requires a stochastic estimate, either following the method proposed in~\cite{Joo:2001bz}, or by using a Gaussian heat-bath and a rational approximation of the exponential. The idea of using the trace of a function of the Dirac operator to simulate the gauge action was first proposed in~\cite{Horvath:2006md}.
\subsection{Renormalisation of fermion fields}
A block renormalisation group transformation of the fermion fields can be defined in terms of blocking matrices $B^{-1}$ and $\overline{B}^{-1}$ (which must be functions of the gauge fields) and $\alpha$ (which must be independent of the gauge fields)
\begin{align}
Z_0 =& \prod_{j=0}^{N_f-1}\int dU \int d\psi_0^{(j)}d\overline{\psi}^{(j)}_0 e^{-S_g[U] - \sum_{i=0}^{N_f - 1}\overline{\psi}^{(i)}_0 D_0^{(i)} \psi_0^{(i)}} 
\times\nonumber\\
& 
\int d\psi_1^{(j)}d\overline{\psi}^{(j)}_1 e^{ -\sum_{i=0}^{N_f - 1}(\overline{\psi}^{(i)}_1 - \overline{\psi}_0^{(i)} (\overline{B}^{(i)})^{-1})\alpha({\psi}_1^{(i)}- (B^{(i)})^{-1} \psi_0^{(i)})}\propto \nonumber\\
 Z_1 = & \prod_{j=0}^{N_f-1}\int dU \int d\psi_1^{(j)}d\overline{\psi}^{(j)}_1 e^{-S'_g[U] - \sum_{i=0}^{N_f - 1}\overline{\psi}^{(i)}_1 D_1^{(i)} \psi_1^{(i)} + J} 
\end{align}
For simplicity, I will subsequently suppress the flavour indices $i$ and $j$.  For $\alpha=\infty$, we may write
\begin{align}
\overline{B}D_0 B =& D_1&
J=&\Tr[\log D_1 - \log D_0]\nonumber\\
\overline{\psi}_1 =& \overline{\psi}_0 \overline{B}^{-1}&
\psi_1 =& B^{-1}\psi_0.
\end{align}
A possible choice of blocking is
\begin{align}
\overline{B}^{-1} =& (D_1)^{1/2} D_0^{-1/2};&
B^{-1} = &D_0^{-1/2} (D_1)^{1/2} ,
\end{align}
which leads, if $D_1$ is the lattice overlap operator and $D_0$ the continuum Dirac operator, to the Ginsparg-Wilson relations,
\begin{gather}
\gamma_L D_1 + D_1 \gamma_R = 0,
\end{gather}
with
\begin{gather}
\gamma_{L,R} = \gamma_5\left(1-\frac{D_1^{\dagger}D_1}{4}\right)^{1/2} \pm \gamma_5(D_1 - D_1^{\dagger})\left(1-\frac{D_1^{\dagger}D_1}{4}\right)^{-1/2},
\end{gather}
and an associated $\mathcal{CP}$-symmetric lattice chiral symmetry.

In~\cite{Cundy:2009ab}, I introduced a parameter $\zeta$ to control the `latticeness' of the fermion fields. The Dirac operator $D_1$ was re-written in terms of the exponentials, $\zeta^4 e^{-\zeta\sum_{\mu}|x_{\mu}-n_{\mu}|}$, where $n$ is a lattice site and $x$ a position in the continuum, so that as $\zeta\rightarrow\infty$ these exponentials become Dirac $\delta$-functions and a lattice theory is reproduced. While it is difficult to visualise the theory at the target of $\zeta=\infty$, at finite $\zeta$ everything remains well-defined. I showed that, for overlap fermions, the blocking remains valid as the limit $\zeta\rightarrow\infty$ is taken. A full discussion and proof can be found in~\cite{Cundy:2009ab}.

In this work, I consider a similar blocking from a fine lattice overlap operator, $D_0$,  to a coarse lattice overlap operator, $D_1$, each with an underlying fine gauge field $U_0$ and $U_1$ respectively.  I write the blocking, the Dirac operator, gauge fields and fermion fields in terms of a parameter $\xi$, which I shall use to interpolate from the coarse lattice to the fine lattice. We may write
\begin{gather}
D_{\xi} = B^{\dagger}[U_{\xi},\xi] D_0[U_{\xi}]B[U_{\xi},\xi] 
\end{gather}
At $\xi=0$, $B=1$ and $D$ remains the original fine Dirac operator; at $\xi=1$ $D_{\xi}$ is equivalent to a coarse Dirac operator. The fermion field transformation may be written as
\begin{align}
\psi_{\xi} =& e^{\frac{1}{2}\int_0^{\xi} \frac{d}{d\xi'} \log[D_{\xi'}] d\xi'} \psi_0; &\overline{\psi}_{\xi} =& \overline{\psi}_0 e^{-\frac{1}{2}\int^0_{\xi} \frac{d}{d\xi'} \log[D_{\xi'}] d\xi'} ; 
\end{align} 
and an infinitesimal blocking is
\begin{gather}
\psi_{\xi + \delta\xi} = \psi_{\xi} +\frac{1}{2} \delta{\xi} \frac{d}{d\xi} \log[D_{\xi}[U_{\xi},\xi] ]\psi_{\xi}.
\end{gather} 
Thus the Jacobian for this infinitesimal change in the fermion fields together with the change in the gauge action is  
\begin{gather}
\log \det J - \delta[S_g] = \Tr\delta{\xi}\frac{d}{d{\xi}}\left[ S_{\xi}[D]\right],
\end{gather}
where
$S_{\xi}[D] = \log[D_{\xi} ] -  S_g[D_\xi]$
\subsection{Gauge field flows}\label{sec:2.3}
We may define a differential with respect to the gauge fields as
\begin{align}
\partial_{x,\mu}^a f(U) =& \lim_{t\rightarrow 0 }\frac{d}{d t^a} f(U_{t;x,\mu;y,\nu}); &
U_{t;x,\mu; y,\nu} =& \left \{
\begin{array}{l l}
e^{itT^a}U_{x,\mu} &x =y,\mu=\nu \\
U_{x,\mu} & \text{otherwise}
\end{array}\right.
\end{align}
For a gauge field transformation given by,
\begin{gather}
\frac{d}{d\xi} U_{\xi,x,\mu} = i\Pi_{\xi,x,\mu}[U_{\xi},\xi]  U_{\xi,x,\mu},
\end{gather}
the Jacobian, $J_{U}$, of the infinitesimal transformation satisfies~\cite{Luscher:2009eq}
\begin{gather}
 \Tr\log J_U = i\delta\xi \partial^a_{x,\mu} (\Pi_{\xi,x,\mu}[U_{\xi},\xi]). 
\end{gather}
Thus, the solution to the equation
\begin{gather}
\Tr[i\partial^a_{x,\mu} (\Pi^{a}_{x,\mu}) + i\partial^a_{x,\mu}(S) \Pi^a_{x,\mu} + \partial_{\xi}S ]= C_{\xi},\label{eq:fine}
\end{gather}
where $\partial_{\xi}$ excludes the dependence on $\xi$ within the gauge fields, and $C_{\xi}$ is independent of the gauge fields, completes the renormalisation group link between $D_0$ and $D_1$ (and equally the renormalisation group construction of the overlap action). It can be shown that a solution for $Z_{\xi}$ exists for all $\xi$ as long as $S$ and $\partial_{\xi}S$ are analytic, and integrating over this solution will give the required transformation of the gauge fields. Finding this solution is rather more problematic. However, it is possible to map from the coarse gauge field to the fine gauge field stochastically, and for an importance sampling method, that is good enough.
\section{Multi-grid HMC}\label{sec:3}
The detailed balance equation reads,
\begin{gather}
P[U_0^{0}\leftarrow U_0^{1}]W_C[0,U_0^{1}] = P[U_0^{1}\leftarrow U_0^{0}] W_C[0,U_0^{0}] = P[U_0^{1}\leftarrow U_1^{1}]P[U_1^{1}\leftarrow U_1^{0}]P[U_1^{0}\leftarrow U_0^{0}] W_C[0,U_0^{1}],
\end{gather}
where, in $U_{\xi}^{\tau}$, the subscript refers to $\xi$ and the superscript to Monte-Carlo time and  
\begin{gather}
W_C[\xi,U_{\xi}] = \prod_{i=1}^{N_f} \det[D^i[\xi]] e^{-S_g[\xi] }.
\end{gather}
If each of the three steps satisfies detailed balance, then it is clear that the complete update of the gauge field satisfies detailed balance. For the transfer from the fine to the coarse lattice, I introduce the momentum field $\Pi$ according to the Gaussian distribution $e^{-(\Pi +A_{\xi})^2}$, where 
\begin{gather}
A_{\xi} = (\partial^a_{x,\mu}(S))^{-1}(\partial_{\xi}S)_{x,\mu}. 
\end{gather}
This requires that $ (\partial^a_{x,\mu}(S))$ is invertible (I shall leave the proof until the later work), and that the Dirac operator is constructed so that $A_{\xi}$ is independent of the gauge and pseudo-fermion fields at $\xi=0$ and $\xi = 1$. We then proceed by integrating along the equations of motion using a reversible, area conserving procedure, such as the Omelyn integrator, along a trajectory $T$,
\begin{align}
\frac{d}{d\xi} U =& i (\Pi + A_{\xi}) U; &
\frac{d}{d\xi}\Pi =& \partial^a_{x,\mu} S[D_{\xi}] - \frac{d}{d\xi} A_{\xi},
\end{align}
which conserves $E=(\Pi_{\xi}+A_{\xi})^2 + Tr S[D_{\xi}]$. $\Pi$ now is a conjugate field and does not depend on the gauge field. The Jacobian from the gauge field update cancels the Jacobian from the momentum update. Reversibility can be maintained using an iterative procedure. I estimate the trace using pseudo-fermions generated according to a heatbath and a rational approximation $R[D_{\xi}] \sim e^{S[D_{\xi}]}$ obtained using the Remez algorithm, so that $\int d\phi d\phi^{\dagger} e^{-\phi^{\dagger} R \phi} = \det R^{-1} = e^{-\Tr S[D]}$. 
The detailed balance equation for the transformation from $[U_1]$ to $[U_0]$ reads (if the integration is exact)
\begin{align}
P[U_0\leftarrow U_1] W_C[U_1]=& \int d \Pi_1   d U_1  \int d\phi d\phi^{\dagger} e^{-\frac{1}{2}(\Pi_1+A_1)^2}e^{ -\phi^{\dagger}R[D_{\xi}] \phi} \delta([U_0,\Pi_0] - T[U_1,\Pi_1])\nonumber\\
=&\int  d\Pi_0 dU_0d\phi d\phi^{\dagger}  e^{-\frac{1}{2}(\Pi_0+A_0)^2}e^{-\phi^{\dagger}R[D_{\xi}] \phi} \delta([U_1,\Pi_1] - T^{-1}[U_0,\Pi_0])\nonumber\\
=& P[U_1\leftarrow U_0] W_C[U_0].\label{eq:db}
\end{align}
We can correct for the error from the numerical integration either by re-weighting or an extrapolation in the integration step-size. Similarly, the integration from $U_0$ to $U_1$ can be shown to satisfy detailed balance, although for practical purposes this step will not be required, because each coarse configuration can be generated from the same thermalised fine configuration. 

We run enough coarse trajectories to gain an independent configuration, and, if required,  generate a fine configuration using the procedure above. For many quantities, the gauge field associated with the fine Dirac operator will not be needed: calculations can proceed with the coarse Dirac operator and non-perturbative renormalisation will take care of the rest. Only if the finer lattice resolution is needed for a particular observable will it be necessary to perform the costly conversion to the fine lattice and the equally costly calculation of the observable.
\section{Numerical implementation} \label{sec:4}
The initial kernel operator, $K_0$, is defined as a Wilson operator on the fine gauge field:
\begin{gather}
(K_0)_{x,x'} = \gamma_5\left(\frac{1}{\kappa} - \sum_{\mu}\left[(1- \gamma_{\mu})\delta_{x+\hat{\mu},x'}U_{\mu}(x) + (1+ \gamma_{\mu})\delta_{x-\hat{\mu},x'}U^{\dagger}_{\mu}(x-\hat{\mu})\right]\right).
\end{gather}
Using a scale factor of three, I define a blocking operator, $B_K^i$ for the kernel in terms of the parameter $\xi$ and additional parameters $C^i_{\mathfrak{L}}$ (which can be tuned to minimize the force and ensure stability when constructing  the fine gauge),
\begin{align}
(B_K^i)_{y,y'}[\xi]= \sum_n \sum_{\mathfrak{L}_{yy'}} U_{\mathfrak{L}_{yy'}}  [(1-\xi^2)^2 \delta_{yy'} +& (1-(1-\xi^2)^2)\delta_{yn}] C^i_{\mathfrak{L}_{yy'}} \nonumber\\
&\prod_{\beta}\theta\left(\frac{3}{2}-|y_{\beta}-n_{\beta}|\right)\theta\left(\frac{3}{2}-|y'_{\beta}-n_{\beta}|\right),
\end{align}
where  $n_{\mu} \in 1,4,7,\ldots$, $\mathfrak{L}_{xn}$ a path of links between $x$ and $n$ (constrained within the hypercube), and $U_{\mathfrak{L}}$ the corresponding path ordered product of links. For a coarse kernel operator $K_1 = (B^{0}_K)^{\dagger}K_0 B^{0}_K$ the Fourier transform (in the free theory) takes the form $i\gamma_{\mu}\sin(p_{\mu} + (1-\cos(p_{\mu}))+ \iota i\gamma_{\mu}\sin(3 p_{\mu} + (1-\cos(3 p_{\mu}))  + m$ for some $\iota$. For $\iota \ge 1$ this introduces fermion doublers. I have chosen to avoid these doublers by including a second term with different coefficients, so that
\begin{gather}
K_{\xi} = (B_K^0)^{\dagger}K_0 B_K^0 - \alpha (B_K^1)^{\dagger}K_0 B_K^1 + \beta (B_K^0)^{\dagger} B_K^0 - \beta' (B_K^1)^{\dagger} B_K^1 - \gamma (1-(1-\xi^2)^2)(1-\delta_{x,n}\delta_{n,x'}),
\end{gather}
where the parameters $\alpha$, $\beta$, $\beta'$ and $\gamma$ need to be tuned to avoid doublers. A badly tuned set of parameters is usually obvious from the eigenvalue spectrum of the kernel and overlap operators. $B_K$ is constructed so that $\partial_{\xi} B_K = 0$ at $\xi = 0$ and $\xi = 1$, as required by the algorithm described in the previous section. The overlap operator at mass $\mu$ is
$D[\xi] = (1+\mu) + (1-\mu)\gamma_5\sign(K_{\xi}).$

Construction of the HMC algorithm for the coarse Dirac operator is trivial. I am currently testing it on a small  PC-Cluster, using a $4^38$ coarse lattice to generate $12^324$ gauge field ensembles, and comparing it against a $12^24$ trajectory generated using normal methods; timing the coarse multi-grid HMC over trajectories of length 1 and the non-multi-grid comparison over trajectories of length 0.02 and extrapolating to a length 1 trajectory. The multi-grid code gave, on average, a factor of 107 gain over the original routine. There was, however, an  overhead of a factor of three from a normal $4^38$ run, which was mostly caused because both the overlap and Kernel operators were very poorly conditioned compared to the small lattice counterpart, and partly from the additional work needed to differentiate the gauge fields. Obviously, the cost of constructing the fine lattice has yet to be included in this estimate. It is to be hoped that with proper tuning of the kernel operator $K_{\xi}$ that this cost will be no more than that of a single length 1 trajectory on the larger lattice, meaning that overall the gain will be roughly proportional to the autocorrelation length. 
\section{Discussion}\label{sec:5}
I have discussed whether it is possible, using overlap fermions, to generate fine gauge fields ensembles using a coarse Ginsparg-Wilson Dirac operator. The method will not be valid for other lattice Dirac operators because the underlying renormalisation group theory breaks down. Small scale tests show that this method gives a two orders of magnitude gain over my previous HMC algorithm, although with additional factors not yet included in that cost; the most important of which is the cost of extracting physics from the gauge field ensembles. It may reasonably be asked what use is there in generating the ensembles quickly using this method if one then has to apply an overlap operator on an exceptionally large lattice to extract physics. For most quantities it will also be possible to do the physics on the coarse lattice: the coarse and fine lattice observables will be linked to each other by renormalisation, and the renormalisation constants can be calculated non-perturbatively as usual. The number of observables which require the fine lattice will, hopefully, be quite small. It should also be possible to adapt this method to allow an anisotropic formulation on the fine lattice. Why then use this method at all, since it gives a considerable overhead compared to the usual approach of a coarse Dirac operator on coarse gauge field? Because the locality of the overlap operator is improved (which I have confirmed numerically), it will (if correctly constructed) have smaller lattice artefacts, and it will have a greater sensitivity to the topology of the fine gauge field, and, of course, the fine gauge field and larger lattice is available for those observables which require it. 

However, the question of how much of a gain, if any, is achievable by this approach is not really answerable until it has been fully tested. I hope to present numerical results in the subsequent paper.
\section*{Acknowlegments}
I am grateful to the support of the Deutsche Forschungsgemeinschaft grant DFG FOR-465, and for many useful discussions with Andreas Sch\"afer. I am also grateful to the hospitality of the Kivali Institute of Theoretical Physics in Beijing during the workshop ``Lattice Quantum Chromodynamics'' in July 2009. Numerical calculations were performed on the PC Cluster Juli at the J\"ulich Supercomputer Center in Germany.

\bibliographystyle{elsarticle-num-mcite}
\bibliography{proceedings}

\end{document}